\documentclass[twocolumn,showpacs,showkeys,prd]{revtex4}
\usepackage{epsf} \usepackage{color}
\usepackage{amsmath,amsthm,amssymb,amsfonts,amsbsy}
\usepackage{graphicx}
\usepackage{dcolumn}
\usepackage{bm}
\everymath{\displaystyle} 
\begin{document} 

\title{Potential for a new muon \emph{g}-2 experiment}

\author{Alexander J. Silenko}
\email{silenko@inp.minsk.by} \affiliation{Research Institute of
Nuclear Problems, Belarusian State University, Minsk 220030,
Belarus}

\date{\today}

\begin {abstract}
A new high-precision experiment to measure the muon \emph{g}-2
factor is proposed. The developed experiment can be performed on
an ordinary storage ring with a noncontinuous and nonuniform
field. When the total length of straight sections of the ring is
appropriate, the spin rotation frequency becomes almost
independent of the particle momentum. In this case, a
high-precision measurement of an average magnetic field can be
carried out with polarized proton beams. A muon beam energy can be
arbitrary. Possibilities to avoid a betatron resonance are
analyzed and corrections to the \emph{g}-2 frequency are
considered.
\end{abstract}
 \maketitle

\section{Introduction}

Measurement of the anomalous magnetic moment of the muon is very
important because it can in principle bring a discovery of new
physics. Experimental data dominated by the BNL E821 experiment,
$a_{\mu\pm}^{exp}=116592080(63)\times10^{-11}~(0.54$ ppm), are not
consistent with the theoretical result,
$a_{\mu\pm}^{the}=116591790(65)\times10^{-11}$, where $a=(g-2)/2$.
The discrepancy is 3.2$\sigma$:
$a_{\mu\pm}^{exp}-a_{\mu\pm}^{the}=+290(90)\times10^{-11}$
\cite{JandN}. In this situation, the existence of the
inconsistency should be confirmed by new experiments. The past BNL
E821 experiment \cite{finalrept} was based on the use of
electrostatic focusing at the ``magic'' beam momentum
$p_m=mc/\sqrt{a} ~(\gamma_m=\sqrt{1+1/a}\approx 29.3)$. An
upgraded (but not started up) experiment, E969 \cite{Fermilab},
with goals of $\sigma_{syst} = 0.14$ ppm and $\sigma_{stat} =
0.20$ ppm is based on the same principle.

Since the muon \emph{g}-2 experiment is very important, a search
for new methods of its performing is necessary. One of new methods
has been proposed by Farley \cite{FF}. Its main distinctions from
the usual \emph{g}-2 experiments are \emph{i}) noncontinuous
magnetic field which is uniform into circular sectors, \emph{ii})
edge focusing, and \emph{iii}) measurement of an average magnetic
field with polarized proton beams instead of protons at rest. A
chosen energy of muons can be different from the ``magic'' energy.
Its increasing prolongs the lab lifetime of muons. As a result, a
measurement of muon \emph{g}-2 at the level of 0.03 ppm appears
feasible \cite{FF}.

In the present work, we develop the ideas by Farley. We adopt his
propositions to measure the average magnetic field with polarized
proton beams and to use a ring with a noncontinuous field for
keeping the independence of the spin rotation frequency from the
particle momentum. We also investigate the most interesting case
when the beam energy can be arbitrary. However, we propose to
perform the high-precision muon \emph{g}-2 experiment on an
ordinary storage ring with a nonuniform field created by
superconducting magnets. We prove that the independence of the
spin rotation frequency from the particle momentum can be reached
not only in a continuous uniform magnetic field
\cite{finalrept,Fermilab} and a noncontinuous and locally uniform
one \cite{FF} but also in a usual storage ring with a
noncontinuous and nonuniform magnetic field. In the last case, the
total length of straight sections of the ring should be
appropriate. We also analyze possibilities to avoid the betatron
resonance $\nu_x=1$ ($\nu_x$ is the horizontal tune) and consider
corrections to the \emph{g}-2 frequency.

The system of units $\hbar=c=1$ is used.

\section{$\bm g$-2 ring with a
noncontinuous magnetic field and magnetic focusing}

Let us consider spin dynamics in a usual storage ring with a
noncontinuous magnetic field and magnetic focusing. The general
equation for the angular velocity of spin precession in the
cylindrical coordinates is given by (see Ref. \cite{PhysRevST})
\begin{eqnarray}
\bm\omega^{(a)}=-\frac{e}{m}\Biggl\{a\bm B-
\frac{a\gamma}{\gamma+1}\bm\beta(\bm\beta\cdot\bm B)
\nonumber\\
+\frac{1}{\gamma}\left[\bm B_\|
-\frac{1}{\beta^2}\left(\bm\beta\times\bm
E\right)_\|\right]\nonumber\\
\left.+ \frac{\eta}{2}\left(\bm
E-\frac{\gamma}{\gamma+1}\bm\beta(\bm\beta\cdot\bm
E)+\bm\beta\!\times\!\bm B\right)\!\right\},~~~ \bm\beta=\frac{\bm
v}{c}. \label{eqST}\end{eqnarray}

Eq. (\ref{eqST}) is useful for analytical calculations of spin
dynamics with allowance for field misalignments and beam
oscillations. This equation does not contain small terms which can
be neglected. $\eta=4dm/e$ is an analogue of the $g$ factor for
the electric dipole moment, $d$. The sign $\|$ denotes a
horizontal projection for any vector. Thereinafter, the electric
dipole moment will be disregarded. The vertical magnetic field,
$B_z$, 
is the main field in the muon \emph{g}-2 experiment. The spin
precession caused by this field is defined by
\begin{eqnarray}
\omega^{(a)}_z=-\frac{e}{m}aB_z.
\label{eqSTz}\end{eqnarray}

Let $\bm\Omega^{(a)}$ denotes the average value of
$\bm\omega^{(a)}$. The spin coherence is kept when
\begin{eqnarray}
\frac{d\Omega^{(a)}_z}{dp}=0.\label{main}\end{eqnarray} For a
storage ring with a noncontinuous field, the quantity $B_z$ should
be averaged.

This condition defines a spin-isochronous ring, i.e., the spin
precession frequency is independent of the momentum at the first
order.

Condition (\ref{main}) can be satisfied for ordinary storage rings
with magnets creating nonuniform field (Fig. 1). Beam direction is
normal to the magnet faces and there is not edge focusing. The
number of bending sections can be different. If the field created
by the magnets is given by $B_z(\rho)=const\cdot\rho^{-n}$, the
field index and betatron tunes into bending sections are equal to
$$n=-\frac{R_0}{B_0}\left(\frac{\partial
B_z}{\partial\rho}\right)_{\rho=R_0},~~~\nu_x^{(b)}=\sqrt{1-n},~~~
\nu_z^{(b)}=\sqrt{n},$$ where $B_0\equiv B_z(R_0),~x=\rho-R_0$,
and $R_0$ is the ring radius. Average angular frequency of spin
precession is given by
\begin{eqnarray}
\Omega^{(a)}=\frac{\omega^{(a)}_z\pi\rho}{\pi\rho+L}=-\frac{\pi
ea\rho B_z(\rho)}{m(\pi\rho+L)}, \label{eqSTO}\end{eqnarray} where
$L$ is a half of the total length of the straight sections (Fig.
1). The muon anomaly is equal to
\begin{eqnarray}
a_\mu=\frac{g_p-2}{2}\frac{m_\mu}{m_p}\frac{\Omega^{(a)}_\mu}{\Omega^{(a)}_p},
\label{muan}\end{eqnarray} where the fundamental constants $g_p$
and $m_\mu/m_p$ are measured with a high precision. The magnetic
field is the same for muons and protons when they move on the same
trajectory. In this case, their momenta coincide.

\begin{figure}[htb]
\centering {
\begin{minipage}[htb]{8.6cm}
{\includegraphics[scale=0.48]{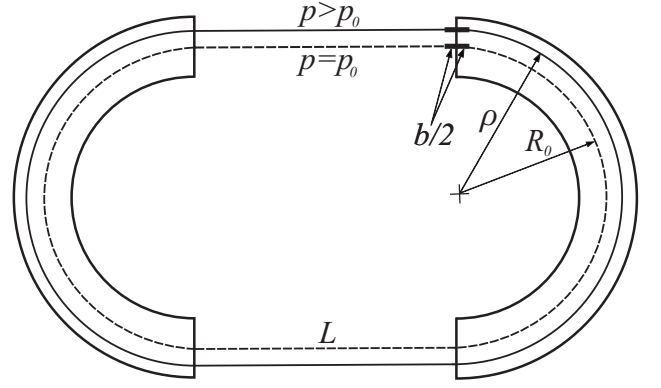}} \caption{The storage
ring.}\label{fig1}
\end{minipage}}
\end{figure}

When the momentum increases ($p>p_0$), the magnetic field becomes
weaker, but the time of flight in the magnetic field becomes
longer. The spin precession is defined by
\begin{eqnarray}\frac{d\Omega^{(a)}_z}{dp}=\frac{d\Omega^{(a)}_z}{d\rho}\left(\frac{dp}{d\rho}\right)^{-1},
~~~\frac{dp}{d\rho}=(1-n)eB_z(\rho).\label{eqOm}\end{eqnarray}

Condition (\ref{main}) leads to $d\Omega^{(a)}_z/d\rho=0$ and is
satisfied when
\begin{eqnarray}
L=L_0=\frac{n}{1-n}\pi R_0, \label{eqL}\end{eqnarray} where $R_0$
corresponds to $p_0$ and $B_0$. In this case
\begin{eqnarray}
R_0=\frac{p_0}{|e|B_0}, ~~~
\Omega^{(a)}=\Omega^{(a)}_0=-(1-n)\frac{eaB_0}{m},
\label{eqO}\end{eqnarray} and the following relation takes place:
$$ 
\frac{\Delta C}{C_0}=\frac{\Delta p}{p_0}=(1-n)\frac{x}{R_0},
$$
where $C$ is the orbit circumference. As a result, the momentum
compaction factor is
\begin{eqnarray} \alpha=\frac{\Delta C/C_0}{\Delta p/p_0}=1.
\label{mcf}\end{eqnarray} Since $$\frac{\Delta
C}{C_0}=\frac{1}{\gamma_0^2}\frac{\Delta p}{p_0}+\frac{\Delta
T}{T_0},$$ where $T$ is the revolution period, the definition of
$\alpha$ can be brought to the usual form:
$$\frac{\Delta
T}{T_0}=\left(\alpha-\frac{1}{\gamma_0^2}\right)\frac{\Delta
p}{p_0}.$$ Evidently, the spin-isochronous ring ($\alpha=1$) is
not isochronous in the usual sense, i.e., the beam revolution
frequency depends on the momentum.

Eq. (\ref{eqL}) is not exact because it does not include a
correction for the fringe field. This field also contributes to
the average field, but it is independent of $\rho$. The fringe
field is important only near the magnet edges and causes the
correction to $L_0$ of order of the ratio of the magnet gap to the
ring radius ($\sim10^{-2}$). This correction depends on the number
of the straight sections and can be analytically and numerically
calculated because the magnet field is known with a needed
accuracy.

Evidently, the correction to the local value of $\omega^{(a)}_z$
is given by
$$\delta\omega^{(a)}_z/\omega^{(a)}_z=\delta B_z/B_z.$$
The corrected values of $L_0$ also coincide for muons and protons
because particles with equal momenta move 
in the same field.

Two other corrections to the angular velocity of spin precession
caused by the longitudinal magnetic field and the vertical
betatron oscillations are considered in Section IV. While these
corrections are different for the muons and protons, they are
rather small ($\sim1$ ppm).

The real value of the length of the straight section, $L$, can
slightly differ from $L_0$. In the general case,
\begin{eqnarray} \alpha=\frac{\pi R_0}{\pi R_0+L-L_0}.
\label{mcfg}\end{eqnarray} The difference between the real and
nominal values of the average angular frequency of spin rotation
is given by
\begin{eqnarray} \frac{\Omega^{(a)}-\Omega^{(a)}_0}{\Omega^{(a)}_0}=n
\cdot\frac{L-L_0}{L_0}\cdot\frac{p-p_0}{p_0}\nonumber\\-\frac{n}{2(1-n)}
\cdot\frac{(p-p_0)^2}{p_0^2}. \label{DOf}\end{eqnarray}

It is important that Eq. (\ref{DOf}) does not depend explicitly on
$B$. The first term in the r.h.s. of this equation disappears if
we \emph{define} $L_0=L$. In this case, $p_0$ is the vertex of a
parabola in the momentum space. To find $p_0$ and adjust the ring
lattice, one can make measurements with proton beams. Three
measurements with different values of $p$ are sufficient. The
average proton momentum can be kept with radio frequency (RF)
cavities put into straight sections of the ring. The longitudinal
electric field in the cavities does not influence the spin
dynamics.

\section{Avoiding a betatron resonance}

Condition (\ref{main}) leading to Eq. (\ref{mcf}) should not be
exactly satisfied. It can be shown that the relation $\alpha=1$
leads to the betatron resonance $\nu_x=1$ which results in zeroth
frequency of horizontal coherent betatron oscillation (CBO) of the
beam as a whole and a loss of the beam \cite{CourSny}. Therefore,
the total length of the straight sections should slightly differ
from $L_0$ so that the CBO tune would be small but nonzero:
\begin{eqnarray} \nu_{CBO}\!\equiv\!|1-\nu_x|\!=\!\left|1\!-\!\sqrt{1+\lambda}\right|\!\ll\!1,~\lambda=\frac{L-L_0}
{L_0}n.\label{CBOt}\end{eqnarray}

Typically, in a weak focusing ring $\alpha>1$. Eq. (\ref{mcfg})
results in $L<L_0$. We expect that the CBO tune about $0.01$ is
sufficient to keep the beam. In this case, the appropriate choice
of the total length of straight sections $\lambda\sim0.01$ reduces
the dependence of the spin rotation frequency on the beam momentum
by
two orders of magnitude. 
As a result, the use of proton beams for measuring the average
magnetic field becomes quite possible.

Experimental details depend on the beam momentum. If it is higher
than in the completed experiment (see Ref. \cite{FF}), the muon
lifetime in the laboratory frame increases and the RF cavities may
be helpful not only for protons but also for muons to keep the
spin coherence. Otherwise, the use of low muon momentum ($\sim0.3$
GeV/$c$) and much higher statistics (see Ref. \cite{J-PARC}) may
even be more preferable. In this case, the RF cavities are
unnecessary for muons.

\section{Corrections to the $\bm g$-2 frequency}

The problem of taking into account corrections to the 
\emph{g}-2 frequency is very important. One of the main problems
is an influence of the radial and vertical betatron oscillations
on the average vertical magnetic field. We can consider the case
when the velocity of unperturbed motion, $v_0$, coincides with the
absolute value of the velocity of perturbed motion. For the latter
motion, the average longitudinal component of the velocity is
approximately equal to
\begin{eqnarray} v_{\phi}=v_0\left(1-\frac{<v_{\rho}^2+
v_{z}^2>}{2v_0^2}\right)=v_0\left(1-\frac{v_{0\rho}^2+
v_{0z}^2}{4v_0^2}\right).\label{vel}\end{eqnarray}

It can be shown that the average magnetic field for the perturbed
motion, $\overline{B_p}$, slightly differs from that for the
unperturbed motion, $\overline{B_u}$:
\begin{eqnarray} \overline{B_p}=\frac{1-n}{1+\lambda\left(1+\frac{v_{0\rho}^2+
v_{0z}^2}{4v_0^2}\right)}B_0, ~~~ 
\overline{B_u}=\frac{1-n}{1+\lambda}B_0,\label{Bav}\end{eqnarray}
where $\lambda$ is given by Eq. (\ref{CBOt}). Approximately,
\begin{eqnarray} \overline{B_p}=\left(1-\frac{\lambda}{1+\lambda}\cdot\frac{v_{0\rho}^2+
v_{0z}^2}{4v_0^2}\right)\overline{B_u}.\label{Bapp}\end{eqnarray}

When $v_{0\rho}/v_0\sim v_{0z}/v_0\sim0.001,~\lambda\sim0.01$, the
correction to the average vertical magnetic field for the betatron
oscillations is rather small and may be even negligible.

A noncontinuous vertical magnetic field leads to a longitudinal
magnetic field on the edges of the magnets. Possibly, the latter
field is a reason of the main correction to the \emph{g}-2
frequency. It was asserted in Ref. \cite{MRR} that this field
causes ``the need to know $\int{\bm B\cdot d\bm l}$ for the muons
to a precision of 10 ppb''. However, we should take into account
that the longitudinal magnetic field cannot be neglected only on
small segments of the beam trajectory near edges of magnets. As a
result, the above estimate of precision should be decreased by
several orders of magnitude.

The correction for the longitudinal magnetic field can be
carefully examined. As ${\rm curl}\,\bm B=0$ and
$B_\phi=(z/\rho)(\partial B_z/\partial \phi)$, the longitudinal
magnetic field acting on a particle oscillates. When the vertical
velocity oscillation (pitch) is given by
$v_z/v_0=\psi_0\cos{(\omega_vt+\delta)}$,
\begin{eqnarray} B_\phi=z_0\frac{\partial B_z}{\partial l}\sin{(\omega_vt+\delta)}, ~
z_0=
\frac{\psi_0\omega^{(b)}_cR_0}{\omega_v}=\frac{\psi_0R_0}{\sqrt{n}},\label{Bphi}\end{eqnarray}
where $l$ is the trajectory length and $\omega^{(b)}_c=v_0/R_0$ is
the cyclotron frequency into bending sections. Evidently,
$\int{B_\phi dl}=z_0B_0\sin{(\omega_vt+\delta)}$.

To estimate the correction, we can suppose that $(\partial
B_z)/(\partial l)\approx B_0/b$. The length of the considered
trajectory segment at the magnet edge is $b$. Calculations can be
simplified if we present the angular velocity of the spin
precession in the form
\begin{eqnarray} \bm\omega_a=a_0\bm e_z+a_2\sin{(\omega_vt+\delta)}\bm e_\phi, \nonumber\\
a_2=-\frac{eB_0}{m}\cdot\frac{(a+1)\psi_0R_0}{\sqrt{n}\gamma
b}\label{omega}\end{eqnarray} and suppose that $a_0\approx
-eB_0a/(2m)=const$. This is nothing but an estimate because the
vertical magnetic field strongly varies within the considered
trajectory segment.

To calculate the correction, we can use the results presented in
Ref. \cite{PhysRevST}. The corrected local angular frequency
is given by
\begin{eqnarray} \omega_z^{(l)}=a_0\left[1-\frac{a_2^2}
{4(\omega_v^2-a_0^2)}\right].\label{claf}\end{eqnarray}

The correction to the average angular velocity of the spin
precession caused by the longitudinal magnetic field is equal to
\begin{eqnarray}
\delta\Omega_{l}^{(a)}=\frac{\omega^{(a)}_z(2\pi\rho-b/2)
+\omega_z^{(l)}b}{(2\pi\rho-b/2)+b+(2L-b/2)}-\Omega^{(a)}
\nonumber\\
=-\Omega^{(a)}\frac{a_2^2}
{2\left[4\omega_v^2-(\omega^{(a)}_z)^2\right]}\cdot\frac{b}{2\pi\rho}
\nonumber\\
\approx -\Omega^{(a)}\frac{(a+1)^2\psi_0^2R_0} {4\pi
n(4n-a^2\gamma^2)b}, \label{eqSTC}\end{eqnarray} where
$\omega^{(a)}_z$ and $\Omega^{(a)}$ are given by Eqs.
(\ref{eqSTz}) and (\ref{eqSTO}), respectively.

Eq. (\ref{eqSTC}) defines only the correction for one segment of
the beam trajectory. To obtain the total correction, we should
take into account the Maxwell equation $\oint{\bm B\cdot d\bm
l}=0$. If the \emph{g}-2 precession did not take place, the total
effect of the longitudinal magnetic field would vanish. However,
the total correction is nonzero owing to a non-commutativity of
rotations and is provided by the spin component orthogonal to the
beam polarization at the beginning of a beam turn. Therefore, the
total correction, $\Delta\Omega_{l}^{(a)}$, can be obtained with
the multiplication of $\delta\Omega_{l}^{(a)}$ by the additional
factor:
\begin{eqnarray}
\Delta\Omega_{l}^{(a)}\sim F\cdot\delta\Omega_{l}^{(a)}, ~
F\!=\!\left\{\begin{array}{c} \omega^{(a)}_z/\omega^{(b)}_c\!=\!a\gamma
~{\rm when}~ a\gamma\!<\!1 \\ 1~{\rm when}~ a\gamma\geq1
\end{array}\right.
\label{eqSTf}\end{eqnarray}

The quantity $b$ is usually of the order of the magnet gap. If we
substitute the parameters of the BNL E821 experiment into Eqs.
(\ref{eqSTC}),(\ref{eqSTf}), we obtain
$|\Delta\Omega_{l}^{(a)}/\Omega^{(a)}|\sim1$ ppm for both muons
and protons.

To measure the total correction with an absolute accuracy of 0.01
ppm, one should determine the magnetic field parameters with a
relative accuracy of $10^{-3}\div10^{-2}$. Since the field of
magnets is well known, extra measurements may be unnecessary. When
the muon beam momentum is significantly decreased as compared with
the BNL E821 experiment (see Ref. \cite{J-PARC}), the correction
for the muons becomes an order of magnitude less. For low-momentum
beams, one can suppress the vertical betatron oscillations and
additionally reduce the corrections for both the muons and
protons.

In the proposed experiment, the correction for the vertical
betatron oscillations (pitch correction) \cite{fff} (see also Ref.
\cite{PhysRevST}) should also be taken into account. Known
formulas \cite{PhysRevST,fff} give the order of magnitude of this
correction ($\sim 0.1\div1$ ppm). The pitch correction can also be
reduced with a suppression of the vertical betatron oscillations
for low-momentum beams. Specific calculations should allow for a
noncontinuity and a nonuniformity of the magnetic field.

In any case, all the corrections can be determined with an
accuracy of 0.01 ppm or even better.

\section{Discussion and summary}

The stabilization and monitoring the magnetic field is an
important and rather difficult problem. To stabilize the magnetic
field in a few minutes needed for measuring the proton spin
precession frequency, superconducting magnets can be used. It is
more difficult to avoid a change of the magnetic field when
switching from muon to proton storage. However, such a change can
be properly determined. The average magnetic field can be
calculated if the beam momentum and the average radius or
frequency of the beam orbit are known. A change of the average
magnetic field brings a corresponding change of the average radius
and frequency of the beam orbit. Therefore, measuring the
frequencies \cite{Fpc} or positions of the muon and proton beam
orbits allows to determine the shift of the average magnetic
field. The average proton momentum is defined by the RF cavities.
In addition to the muon measurements, proton beams before and/or
after muon runs can be used. The use of these methods should
provide a determination of the shift of the average magnetic field
with a relative accuracy of 0.1 ppm or even better. As a result,
the muon and proton measurements can be related with a high
precision.

The methods of measurement of the \emph{g}-2 precession in the
proposed experiment and the Farley's one are very similar. The
important advantage of a noncontinuous nonuniform ring versus a
noncontinuous uniform one is a possibility to avoid much shimming
needed for creating the uniform magnetic field. Shimming is even
more difficult for the noncontinuous uniform ring than for a
continuous uniform one because of the fringe field. We expect that
the proposed experiment can be carried out with one of existing
rings.

The systematical errors considered above do not prevent to measure
the muon \emph{g}-2 factor with a high precision. The sum of all
systematical errors considered in the manuscript causes less
systematic uncertainty than that in the planned E969 experiment
\cite{Fermilab}. While there are many other systematical errors,
we expect that the precision of the proposed experiment may be
approximately the same or 
better than that of the planned E969 experiment.

A more 
detailed theoretical analysis should be based on the matrix
method. The use of the matrix method is necessary for further
theoretical investigations.
However, any 
theoretical analysis is not sufficient to calculate the spin
dynamics in specific \emph{g}-2 rings with a needed accuracy.
Nevertheless, necessary calculations can be carried out with spin
tracking.

Since the theoretical predictions and the experimental data do not
agree, performing new experiments based on different ring lattices
is necessary. Such experiments will be very important even if they
will not provide better precision as compared with the usual
\emph{g}-2 experiments \cite{finalrept,Fermilab}.

In this work, we propose the new experiment to measure the muon
\emph{g}-2 factor. The developed experiment does not require much
shimming. This experiment could provide an independent
experimental result with different systematics and the advantages
mentioned in the Farley's paper \cite{FF}.

\section*{Acknowledgments}

The author is very much obliged to F.J.M. Farley for valuable
remarks and discussions. The author is also grateful to I.N.
Meshkov and Y.K. Semertzidis for helpful discussions and the
referees for valuable comments and remarks. The work was supported
by the Belarusian Republican Foundation for Fundamental Research
(Grant No. $\Phi$10D-001).

\end{document}